# Perceiving the crust in 3D: a model integrating geological, geochemical, and geophysical data

**Virginia Strati[1,2], Scott A. Wipperfurth[3], Marica Baldoncini[2,4], William F. McDonough[3,5], Fabio Mantovani[2,4]**

[1]INFN, Legnaro National Laboratories, Padua, Italy;

[2]Department of Physics and Earth Sciences, University of Ferrara, Ferrara, Italy;

[3]Department of Geology, University of Maryland, College Park, Maryland, 20742 USA;

[4]INFN, Ferrara Section, Ferrara, Italy;

[5]Department of Earth and Planetary Materials Science and Research Center for Neutrino Science, Graduate School of Science, Tohoku University, Sendai, Miyagi 980-8578, Japan.

Corresponding author: Virginia Strati (strati@fe.infn.it)

**Key Points:**

- 3-Dimensional crustal model around Sudbury Neutrino Observatory (Canada)
- Prediction of geoneutrino signal and crustal heat production in Sudbury region
- Bivariate analysis of U and Th distribution in sampled geologic units

# Abstract

Regional characterization of the continental crust has classically been performed through either geologic mapping, geochemical sampling, or geophysical surveys. Rarely are these techniques fully integrated, due to limits of data coverage, quality, and/or incompatible datasets. We combine geologic observations, geochemical sampling, and geophysical surveys to create a coherent 3-D geologic model of a 50 × 50 km upper crustal region surrounding the SNOLAB underground physics laboratory in Canada, which includes the Southern Province, the Superior Province, the Sudbury Structure and the Grenville Front Tectonic Zone. Nine representative aggregate units of exposed lithologies are geologically characterized, geophysically constrained, and probed with 109 rock samples supported by compiled geochemical databases. A detailed study of the lognormal distributions of U and Th abundances and of their correlation permits a bivariate analysis for a robust treatment of the uncertainties. A downloadable 3D numerical model of U and Th distribution defines an average heat production of $1.5^{+1.4}_{-0.7}$ µW/m$^3$, and predicts a contribution of $7.7^{+7.7}_{-3.0}$ TNU (a Terrestrial Neutrino Unit is one geoneutrino event per $10^{32}$ target protons per year) out of a crustal geoneutrino signal of $31.1^{+8.0}_{-4.5}$ TNU. The relatively high local crust geoneutrino signal together with its large variability strongly restrict the SNO+ capability of experimentally discriminating among BSE compositional models of the mantle. Future work to constrain the crustal heat production and the geoneutrino signal at SNO+ will be inefficient without more detailed geophysical characterization of the 3D structure of the heterogeneous Huronian Supergroup, which contributes the largest uncertainty to the calculation.

# 1. Introduction

Geoscientists map out and define the surface geology and from that predict 3D cross sections of regional terrains. Geological mapping in 3-D is a fundamental task for understanding the potential for economic resources and the geological evolution of a region. Infrequently are datasets from these surface campaigns fully integrated into a coherent depth projection using data from shallow geophysical surveys. Although geological data of various sorts have been collected almost everywhere on Earth, crustal data in most regions have vastly different resolution and data types that present challenges to integrate into a coherent 3-D picture that projects 10+ km into the crust. With the advent of advanced techniques of statistical analysis and extensive data collection with comparable uncertainties, it is now possible to integrate many different types of information into a single coherent model. The resultant models are useful in geophysical modeling (e.g., structural analysis, geodynamic simulations, seismic wave corrections, and heat flux), geologic interpretation (e.g., orogenic history, past environments, and crustal processes), and particle physics (e.g., geoneutrinos flux and muon tomography).

We report here a method of integrating available geological, geochemical, and geophysical data into a coherent 3-D model of the upper crust of the Sudbury region of Canada (see supporting information **Dataset S1**). Our efforts build on a previous study [*Huang et al.*, 2014], hereafter H14, that developed a 3-D model of the thick LOcal Crust belonging to the 6°× 4° (~440 km × 460 km total area) region centered near Sudbury (hereafter defined as LOC) (**Figure 1**). H14 found that the Huronian Supergroup of the Southern Province was chemically and lithologically heterogeneous and revealed marked variations in its K, Th, and U contents. Consequently, predictions of the abundance and distribution of the heat producing elements in this unit came with considerable uncertainties, resulting in a large variability on estimates of the local radiogenic heat power and expected geoneutrino signal at the SNO+ detector located in

Sudbury. Based on these findings, we performed additional geochemical sampling (112 new analyses) of the region and combined these data with the models published in H14 and in [*Olaniyan et al.*, 2015], to build a revised 3-D high-resolution model that describes the Close Upper Crust (CUC) corresponding to the 50 km × 50 km area around SNO+.

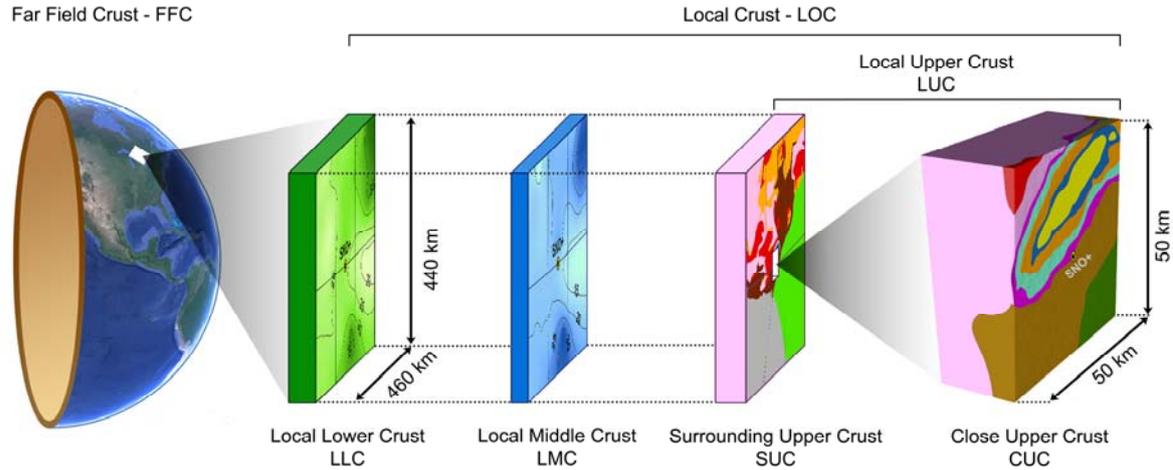

**Figure 1.** The crustal geoneutrino signal expected in SNO+ is calculated considering the Far Field Crust (FFC, the rest of the Earth's crust not included in the studied 6° × 4° region) and the LOcal Crust (LOC, the 6°× 4° regional area under study). Adopting the same structure of H14, the LOC is subdivided in Local Lower Crust (LLC), Local Middle Crust (LMC) and LUC (Local Upper Crust). The latter includes the Surrounding Upper Crust (SUC) and the Close Upper Crust (CUC), i.e. the closest 50 x 50 km region investigated in this study.

## 2. Motivation

Motivation of H14 and this study was to build a model that would then be used to calculate the expected geoneutrino signal at the SNO+ detector, which is a multipurpose kiloton-scale liquid scintillation detector located 2092 (± 6) m underground at SNOLAB outside Sudbury [*Lozza*, 2016; *Sonley*, 2009]. Integrating the 3-D geophysical (i.e. density and spatial distribution of units) and geochemical (i.e. K, Th, and U concentrations) data with the existing surface data yields a more coherent geological understanding of the regional crust surrounding Sudbury.

Geoneutrinos are electron antineutrinos emitted in beta minus decays, with those occurring along the $^{238}$U and $^{232}$Th decay chains having sufficient energies to be detected [*Araki et al.*, 2005]. One of the challenging goals that the SNO+ experiment wants to address in the geoneutrino field are the separation of $^{238}$U and $^{232}$Th geoneutrino spectral components together with the distinction between the mantle and the crustal contributions in a global analysis of the geoneutrino spectrum, comprising data coming from the ongoing KamLAND [*KamLAND Collaboration*, 2013] and Borexino [*Borexino Collaboration*, 2015] experiments. Insights into the mantle contribution to the geoneutrino signal at any individual detector can be pursued provided precise and accurate knowledge of the dominant geoneutrino background, mostly due to reactor antineutrinos, and a refined regional-scale model of the continental crust [*Baldoncini et al.*, 2015].

Understanding the power inside the Earth that drives plate tectonics, mantle convection, and the geodynamo are fundamental goals in our science. The emerging field of neutrino geoscience provides a new tool by which to define the abundance and distribution of heat producing elements inside the Earth. At any given geoneutrino detector that is sited on continental crust, the mantle contribution is only 20% - 25% of the total signal [see Figure 2 in *Šrámek et al.*, 2016]. Thus, to define the mantle contribution and power of the largely inaccessible Earth, it is crucial to understand the specific attributes of the local crustal contribution to the signal. Importantly, global geoneutrino models provide flux maps for the Earth [*Usman et al.*, 2015] which will be a reference for discriminating among distinct compositional paradigms of the bulk silicate Earth [*Dye*, 2010; *Fiorentini et al.*, 2007; *Šrámek et al.*, 2016].

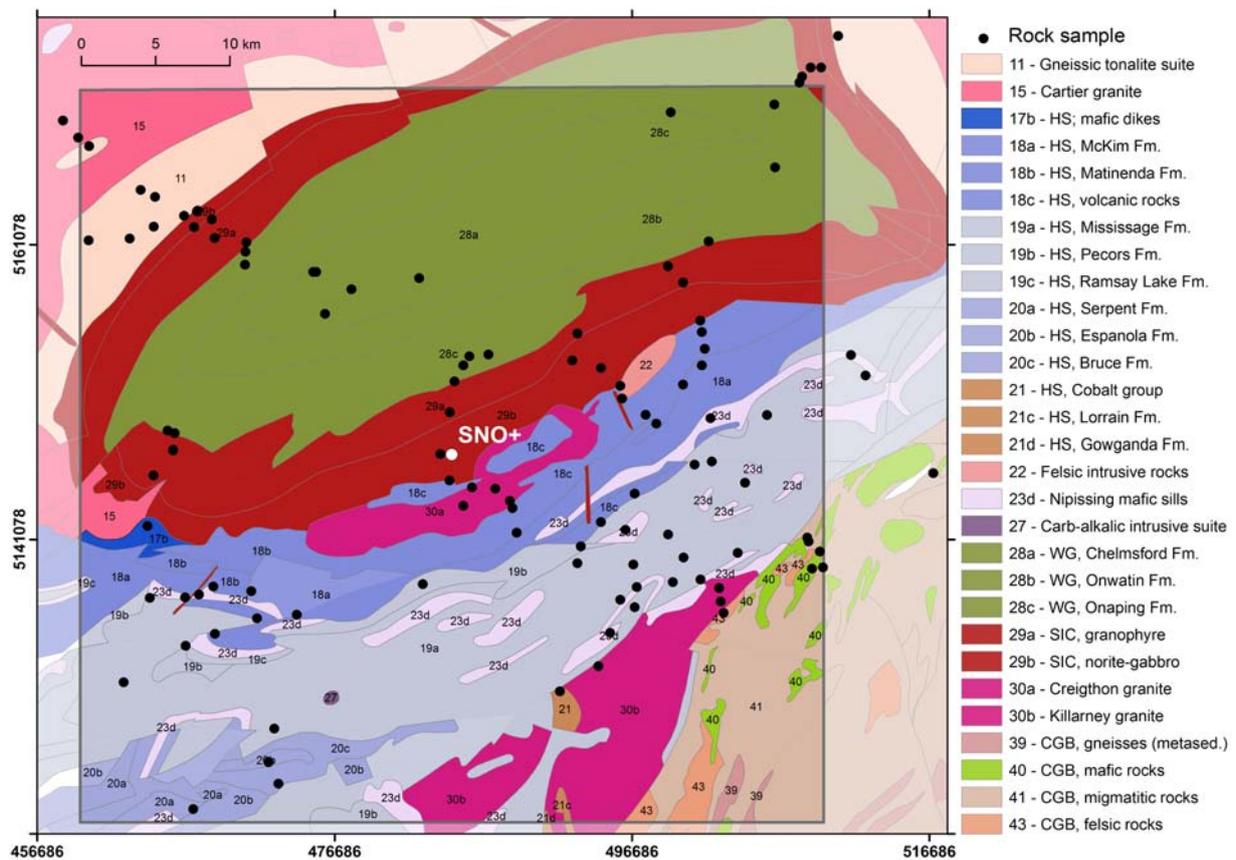

**Figure 2.** Location of the 112 rock samples. Rocks samples are collected in the CUC (inner box) and projected onto the Bedrock Geology of Ontario [Ontario Geological Survey, 2011] (HS = Huronian Supergroup, WG = Whitewater Group, SIC = Sudbury Igneous Complex, CGB = Central Gneiss Belt). (Cartographic reference system NAD1927 UTM Zone 17N).

## 3. Geological setting

The Close Upper Crust (CUC), i.e. the 50 × 50 km region centered at SNOLAB, is the target area of the 3D crustal model constructed for estimating the geoneutrino signal at SNO+.

The study area is comprised mostly of the Southern Province and Sudbury Structure, and lesser areas of the Superior Province and the Grenville Front Tectonic Zone (GFTZ).

The Southern Province, covering much of the southwestern part of the study area, is primarily composed of Huronian Supergroup (HS), a well-exposed Paleoproterozoic succession deposited between 2.4 and 2.2 Ga as the result of a partial Wilson cycle with the rifting and development of a southward-facing passive margin [*Young et al.*, 2001]. The HS can reach up to 12 km of thickness and it is composed of (from bottom to top) the Elliot Lake, Hough Lake, Quirke Lake, and Cobalt groups. A generalized stratigraphic column of the formations of HS is reported in Figure 5 of [*Young*, 2013]. The different groups include variable lithologies, such as sandstones, mudstones, carbonates, conglomerates, and minor volcanic rocks [*Long*, 2004; 2009]. In the study area, the HS is represented primarily by the Elliot Lake Group, a thick package of volcanic rocks and deep-water sediments, and the Hough Lake group, a basal diamictite that fines upward from mudstone to sandstone. The upper formation of the Hough Lake Group, the Mississage Fm., representing 18% of the total studied area, is made up of medium to coarse grained, arkosic to subarkosic sandstones. In the southwest area, carbonate rocks of the Quirke Lake group outcrop in a relative small portion of the study area, while the Cobalt Group is almost absent. The supracrustal rocks of the HS are intruded by the mafic dikes and sills of the Nipssing Gabbro, which are less than 100 m thick, and by felsic intrusions, mainly the granitic rocks of the Creighton and Murray plutons [*Riller*, 2009].

Following the HS deposition, a meteorite impact (1.85 Ga) [*Therriault et al.*, 2002] caused the formation of the Sudbury Igneous Complex (SIC) that intrudes the HS and that, together with the Whitewater group, constitutes the Sudbury Structure. The SIC is geographically divided into North, East, and South ranges and the main mass is composed of norite, quartz-gabbro, and granophyre. The basin of the impact crater was later filled by the Whitewater Group sediments, a 2900 m thick assemblage of breccias, hypabyssal intrusions, carbonaceous sediments, and turbidity sequences [*Rousell and Card*, 2009].

In the northwestern part of the studied region are the Archean crystalline rocks of the Superior Province, the Levack Gneiss Complex. These high-grade rocks (tonalite-granodiorite orthogneiss) form a collar, 0.5-5 km wide, around the North and East margin of the SIC. The complex is intruded by the felsic plutonic rocks of the Cartier Batholith [*Rousell and Card*, 2009].

In the southeast corner of the studied area are Grenville Province rocks in a crustal scale shear zone (GFTZ) that mark the northwest edge of the Grenville Orogeny. It is interpreted as a metamorphic transition comprising gneissic and migmatitic rocks originating from HS sedimentary rocks and Nipissing Gabbro that underwent deep metamorphic and granitization processes [*Davidson*, 1997; *Easton*, 2016].

## 4. Sampling survey

Locations of the 112 collected rock samples are reported in **Figure 2** (see supporting information **Table S4**) and are projected on the published 1:250,000 scale Bedrock Geology of Ontario [*Ontario Geological Survey*, 2011] used as a guide for the survey. Sample GPS location and geological information (e.g. geological formation, lithology granulometry, recognized minerals) were recorded. Every sample was collected from fresh outcrops, representative of the geological formation, and placed in a polyethylene bag (**Figure 3a**). Later each sample was crushed, sealed in a polycarbonate container (**Figure 3b**) and left undisturbed for at least 5 weeks with the objective of establishing radioactive equilibrium between $^{226}$Ra and $^{222}$Rn (see Figure 2 of [*Xhixha et al.*, 2016]).

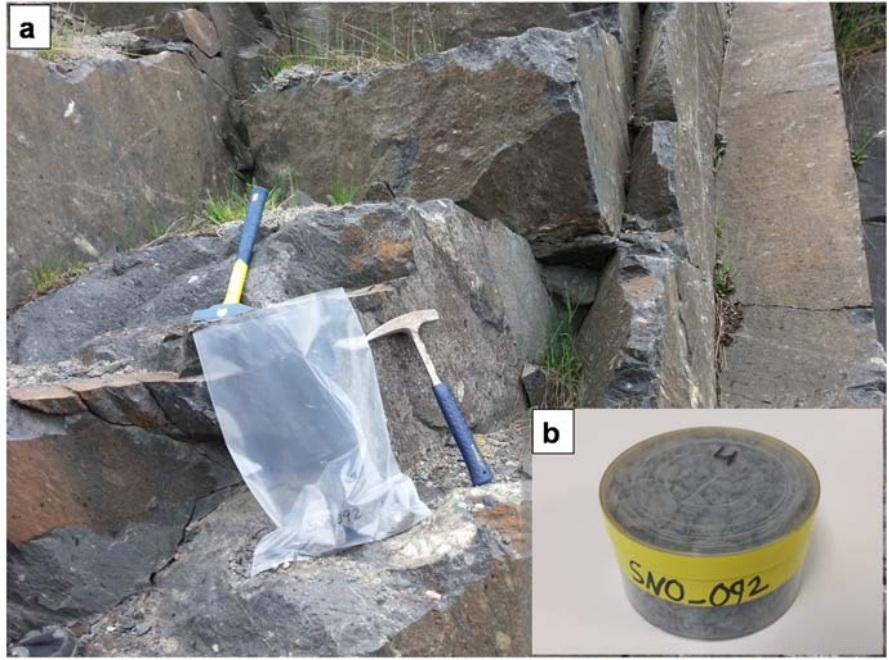

**Figure 3.** Rock sample of lapilli tuff (Geocode 28c, Onaping Fm.). (a) Each sample was collected from fresh outcrop and (b) then crushed and sealed in polycarbonate box of 180 cm$^3$ of volume

Provided the accessibility of the outcrops, the number of the samples collected for each cartographic unit was planned on the basis of the exposure area and the estimated volume, taking into account also the proximity to the detector. For each of the 22 cartographic units, identified by a Geocode, we report extent area, number of samples collected, and average U and Th abundances, with the average ratio between extent area and number of samples being ~15 km$^2$/sample (**Table 1**).

In the CUC area are also homogeneously distributed olivine diabase dikes emplaced along faults cutting across the Sudbury Structure having negligible volumes [*Tschirhart and Morris*, 2012]. Although the reference geological map does not report the presence of the dike swarm according to its spatial resolution, we chose to collect three samples in order to characterize these rocks. After checking that the U and Th abundances of these three samples (**Table 1**) are compatible with the average abundances of the CUC, we decided to exclude them for the geochemical modeling (see Section 6.3) performed with the remaining 109 samples out of the collected 112.

**Table 1.** Summary of the Geocode units, aerial extent, number of sample (N), and average and uncertainties of element abundances. Geocodes of the reference geological map are reported for the corresponding unit abbreviation and the area of the exposed surface. For Geocodes with more than 10 samples the central value and the uncertainty for K, U, and Th abundances are derived from a normal or lognormal distribution fit inferred from the Kolmogorov-Smirnov test (see section 6.2); for the other Geocodes we report the mean and the standard deviation. For Geocodes with one sample the uncertainty corresponds to the statistical uncertainty of the HPGe measurement.

| Unit | Geocode | Group and Formation | Area (km$^2$) | Area (%) | N | K ± σ (%) | U ± σ (µg/g) | Th ± σ (µg/g) |
|---|---|---|---|---|---|---|---|---|
| GT | 11 | Gneissic tonalite suite | 91.5 | 3.7 | 9 | 1.21 ± 0.65 | 0.6 ± 0.9 | 2.8 ± 3.5 |
| CT | 15 | Cartier granite | 62.2 | 2.5 | 2 | 4.55 ± 0.28 | 1.8 ± 1.1 | 56.9 ± 27.3 |
| HI | 17b | HS; Mafic and ultramafic intrusive rocks and mafic dikes | 8.2 | 0.3 | 1 | 0.32 ± 0.03 | 2.0 ± 0.2 | 3.2 ± 0.4 |
| HI | 18a | HS; Elliot Lake Group; McKim Fm. | 121.9 | 4.9 | 7 | 1.94 ± 1.00 | 5.0 ± 3.0 | 16.2 ± 8.4 |
| HI | 18c | HS; Elliot Lake Group; volcanic rocks | 125.3 | 5.0 | 6 | 3.30 ± 1.38 | 5.9 ± 3.2 | 23.2 ± 11.4 |
| HI | 19a | HS; Hough Lake Group; Mississage Fm. | 442.4 | 17.7 | 18 | $1.52^{+1.65}_{-0.79}$ | $2.2^{+2.4}_{-1.2}$ | $6.6^{+7.5}_{-3.5}$ |
| HI | 19b | HS; Hough Lake Group; Pecors Fm. | 41.9 | 1.7 | 4 | 1.37 ± 1.49 | 2.9 ± 1.1 | 9.2 ± 4.3 |
| HI | 20a | HS; Quirke Lake Group; Serpent Fm. | 15.5 | 0.6 | 1 | 2.45 ± 0.15 | 0.9 ± 0.1 | 4.2 ± 0.5 |
| HI | 20b | HS; Quirke Lake Group; Espanola Fm. | 32.3 | 1.3 | 3 | 3.48 ± 0.86 | 3.7 ± 0.7 | 12.5 ± 2.1 |
| HI | 21 | HS; Cobalt Group | 4.6 | 0.2 | 1 | 1.59 ± 0.17 | 1.6 ± 0.2 | 3.1 ± 0.4 |
| HI | 23d | Mafic and related intrusive rocks and mafic dikes | 88.0 | 3.5 | 6 | 0.70 ± 0.51 | 0.4 ± 0.3 | 1.7 ± 1.0 |
| HI | 30a | Felsic intrusive rocks | 53.6 | 2.1 | 4 | 4.36 ± 0.25 | 5.9 ± 2.4 | 36.2 ± 6.4 |
| HI | - | Sudbury Dyke Swarm; olivine diabase | - | - | 3 | 0.48 ± 0.03 | 0.9 ± 0.5 | 4.0 ± 2.8 |
| CM | 28a | Whitewater Group; Chelmsford Fm. | 153.1 | 6.1 | 4 | 1.29 ± 0.29 | 1.1 ± 0.1 | 5.1 ± 0.7 |
| OW | 28b | Whitewater Group; Onwatin Fm. | 160.1 | 6.4 | 2 | 2.19 ± 0.35 | 1.1 ± 0.1 | 5.2 ± 1.4 |
| OP | 28c | Whitewater Group; Onaping Fm. | 343.3 | 13.7 | 12 | 1.83 ± 0.95 | 3.1 ± 0.6 | 8.2 ± 1.0 |
| GN | 29a | SIC; granophyre | 241.7 | 9.7 | 10 | 3.18 ± 0.40 | 3.4 ± 0.5 | 15.1 ± 2.4 |
| NG | 29b | SIC; norite-gabbro | 157.9 | 6.3 | 9 | 1.21 ± 0.08 | 1.3 ± 0.8 | 6.8 ± 4.3 |
| GF | 30b | Felsic intrusive rocks | 126.8 | 5.1 | 3 | 4.15 ± 0.24 | 6.8 ± 5.8 | 31.0 ± 17.5 |
| GF | 40 | Mafic rocks | 16.7 | 0.7 | 1 | 0.96 ± 0.06 | 1.3 ± 0.1 | 4.9 ± 0.5 |
| GF | 41 | Migmatitic rocks and gneisses of undetermined protolith | 124.0 | 5.0 | 5 | 2.41 ± 0.14 | 2.7 ± 2.3 | 13.0 ± 12.1 |
| GF | 43 | Felsic igneous rocks | 18.5 | 0.7 | 1 | 2.33 ± 0.14 | 2.9 ± 0.3 | 4.6 ± 0.5 |

## 5. Analytical method

The radioactive content of the collected samples was measured at the Department of Physics and Earth Sciences of the University of Ferrara, with a High Pure Germanium detector (HPGe) called MCA_Rad. Analytical details are given in [*Xhixha et al.*, 2016; *Xhixha et al.*, 2013]. The overall relative uncertainties on the K, eU and eTh (i.e. U and Th assumed in secular equilibrium) are of the order of 10%. In the analyzed dataset less than 4% of the samples have eU and eTh abundances below the Minimum Detectable Activity (MDA) defined in [*Xhixha et*

*al.*, 2013] and corresponding to about 0.2 and 0.7 µg/g, respectively (see supporting information **Table S4**).

Additional analyses of U and Th on 14 of the 112 samples, including those below MDA of MCA_Rad, were done at the Department of Geology at the University of Maryland using an ICPMS (Thermo-Finnigan Element 2) (see supporting information **Table S1**). These results are reported in supporting information, see **Table S2**. Aliquots of the samples used for gamma ray spectroscopy were powdered and analyzed for U and Th concentrations using a Standard Addition method detailed in [*Gaschnig et al.*, 2016]. U and Th concentrations from Standard Addition have average relative uncertainty of 3.5%.

In addition, external calibration analyses using USGS rock standards were conducted for some 36 other elements including Th and U. The abundances of these elements were calculated by comparison to external standards that were dissolved alongside the samples. We calculated the counts-per-second/concentration of the standard(s) using accepted concentrations from GeoReM (Queried 28 March, 2017). These ratios were compared to counts-per-second for each element within a sample to calculate a final concentration (see supporting information **Table S5**). U and Th results from this External Calibration method agree with the Standard Addition method. Uncertainties on the External Calibration analysis are 5% or better following [*Gaschnig et al.*, 2016].

The U and Th abundances of the five samples below the MDA of MCA_Rad are substituted by the values from ICPMS technique, which has a sensitivity better than HPGe investigation. Taking into account the experimental uncertainties for the remaining nine samples we observe an agreement at 2 sigma level and exclude any systematic effect. The dataset of 112 U and Th abundances is therefore composed by 98 and 14 values from the HPGe and ICPMS technique, respectively.

## 6. Construction of the model

The geological units of the 3D model of the Close Upper Crust (CUC) (see supporting information **Dataset S1**) were defined considering the surface exposure described in the published 1:250,000 scale Bedrock Geology of Ontario [*Ontario Geological Survey*, 2011], which is conveniently simplified according to the spatial resolution of the available information about crustal structure. The upper crust is subdivided into nine units (**Figure 4**) on the basis of lithology, metamorphism, tectonic events, and evolutionary history:

1. Chelmsford Fm., Whitewater Group (CM);

2. Onwatin Fm., Whitewater Group (OW);

3. Onaping Fm., Whitewater Group (OP);

4. Granophyre, Sudbury Igneous Complex (GN);

5. Norite-gabbro, Sudbury Igneous Complex (NG);

6. Cartier Granite (CT);

7. Huronian Supergroup and minor felsic and mafic Intrusions (HI);

8. Grenville Front Tectonic Zone rocks (GF) and

9. Gneissic Tonalite suite (GT).

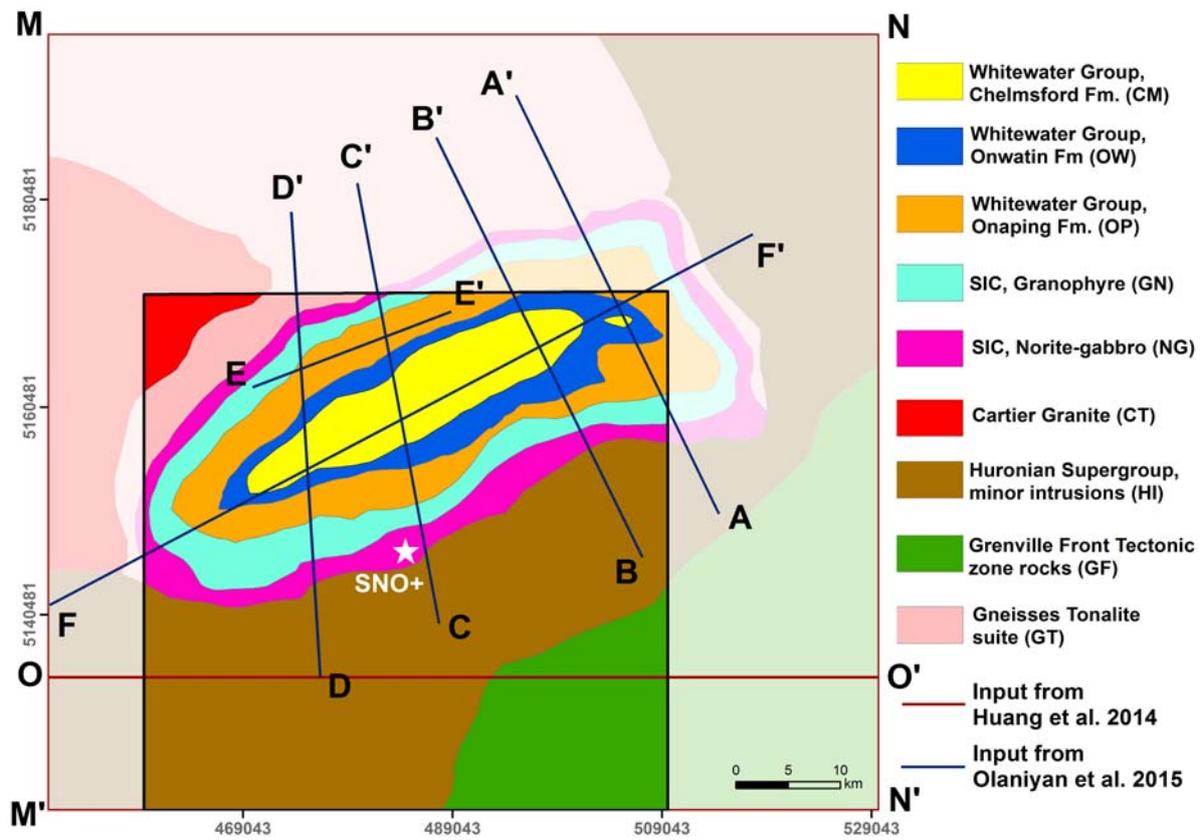

**Figure 4.** Geophysical inputs used for the construction of the 3D model. The six cross sections derived from [Olaniyan et al., 2015] (AA', BB', CC', DD', EE', FF') and the five cross sections extracted from the H14 model (MM', M'N', NN', MN, OO') are projected on the simplified geological map. The inner box represents the CUC. (Cartographic reference system NAD1927 UTM Zone 17N).

The CM, OW, and OP are, respectively the metagraywackes, the pelagic metasedimentary rocks and the breccias of the Whitewater Group that fills the Sudbury Basin while the main mass of the SIC is constituted by granophyre (GN) and norite-gabbro (NG). The HI, formally composed by the HS, includes also minor mafic (Nipissing mafic sills) and felsic intrusions (Creighton and Murray granite). The Gneissic Tonalite suite (GT), that is assumed to be representative of the rest of the upper crust, is an assemblage of high-grade gneissic rocks intruded on the Northwest area by the massive granitic rocks of the Cartier Batholith (CT). In the south-eastern portion of the CUC, the GF unit is characterized by the presence of migmatitic rocks, gneisses and felsic intrusions of the GFTZ. The Geocodes associated to each unit are detailed in **Table 1**.

6.1 Geophysical modeling

The crustal structures of the nine units were defined by combination of multiple geological and geophysical inputs: (i) the contacts of the simplified geological map (Figure 4), (ii) a published digital elevation model [*Jarvis et al.*, 2008], (iii) the map of depth of the top of the middle crust reported in H14, (iv) the 2.5D geological models along six profiles used for constructing the 3D model reported in [Olaniyan et al., 2015] and (v) five virtual cross sections derived from the model developed in H14.

The surface topography for the CUC region uses the digital elevation model produced by the Shuttle Radar Topographic Mission (SRTM) [*Jarvis et al.*, 2008].

The bottom of the 3-D model has a 1×1 km resolution and is the surface of the top of the middle crust (**Figure 1**) determined in H14. The depth map of the top of the middle crust was obtained alongside the error estimation map by applying a geostatistical estimator (Ordinary Kriging) to 343 depth-controlling points. These points are derived from refraction surveys performed in the region surrounding Sudbury. The P-wave velocity of 6.6 km/s is adopted as a contour to identify the top of the middle crust in 18 refraction lines, two of which (XY and AB reported in [*Winardhi and Mereu*, 1997]) are within the CUC area. The top of the middle crust is a 2D input for the construction of the 3D model. The depth of the CUC varies between 16.4 and 20.4 km, with a mean of 18.4 km. The normalized estimation error of the map has an average value of 4.7%.

In [Olaniyan et al., 2015] the 3D model was obtained by integrating a compilation of surface and subsurface geologic data with high-resolution airborne magnetic and gravity data. The authors evaluated qualitatively high resolution Bouguer gravity data with the computed field along with subsurface geologic data and created their cross section profiles. They observed a broad correlation between the measured and computed gravity field and found areas of misfit. The 2.5-D geological models reported in six profiles (AA', BB', CC', CC', EE' and FF' in **Figure 4**), are used as inputs for the modeling of the Sudbury Structure. Orientation data and boundary surfaces of the units of the Whitewater group (CM, OP, and OW units) and of the main mass of the SIC (GN and NG units) are modeled by extracting the depth-controlling points of the boundary surfaces from each profile.

For the remaining area of the CUC, the 3-D geometries of the units were developed in H14 on the basis of surface contacts between units and 16 interpreted crustal cross sections of the area, with the main inputs from [*Easton*, 2000] and [*Adam et al.*, 2000]. In this perspective, five virtual cross sections (MM', NN', MN, M'N', and OO' in **Figure 4**) are extracted from H14 and used as input for inferring the structure of units not constrained by inputs from [*Olaniyan et al.*, 2015].

The geological interfaces of the 9 units are modeled using the interpolator method based on potential field theory [*Calcagno et al.*, 2008] and implemented in the software package GeoModeller. Using the available data from the geological reference map and that reported in [*Olaniyan et al.*, 2015] we reduced interpretational nonuniqueness of the potential field data by applying hard geological constraints, including (i) the stratigraphic succession of geological formations, (ii) geological contacts, (iii) structural data, and (iv) orientation data. **Figure 5** provides 3D views of the determined geological model.

The adopted density values for each unit (**Table 3**) are from the model reported in [*Olaniyan et al.*, 2015] and the relative uncertainties from Table 5 in H14. Density of the HI unit is obtained from the weighted average of values of sediments (2.70 g/cm$^3$) and mafic rocks (2.88 g/cm$^3$), assuming that their proportions are respectively 75% and 25% according to the exposure surface within the reference geologic map. The GT and GF units are assumed to have density equal to the Archean basement value (2.73 g/cm$^3$) reported in [*Olaniyan et al.*, 2015].

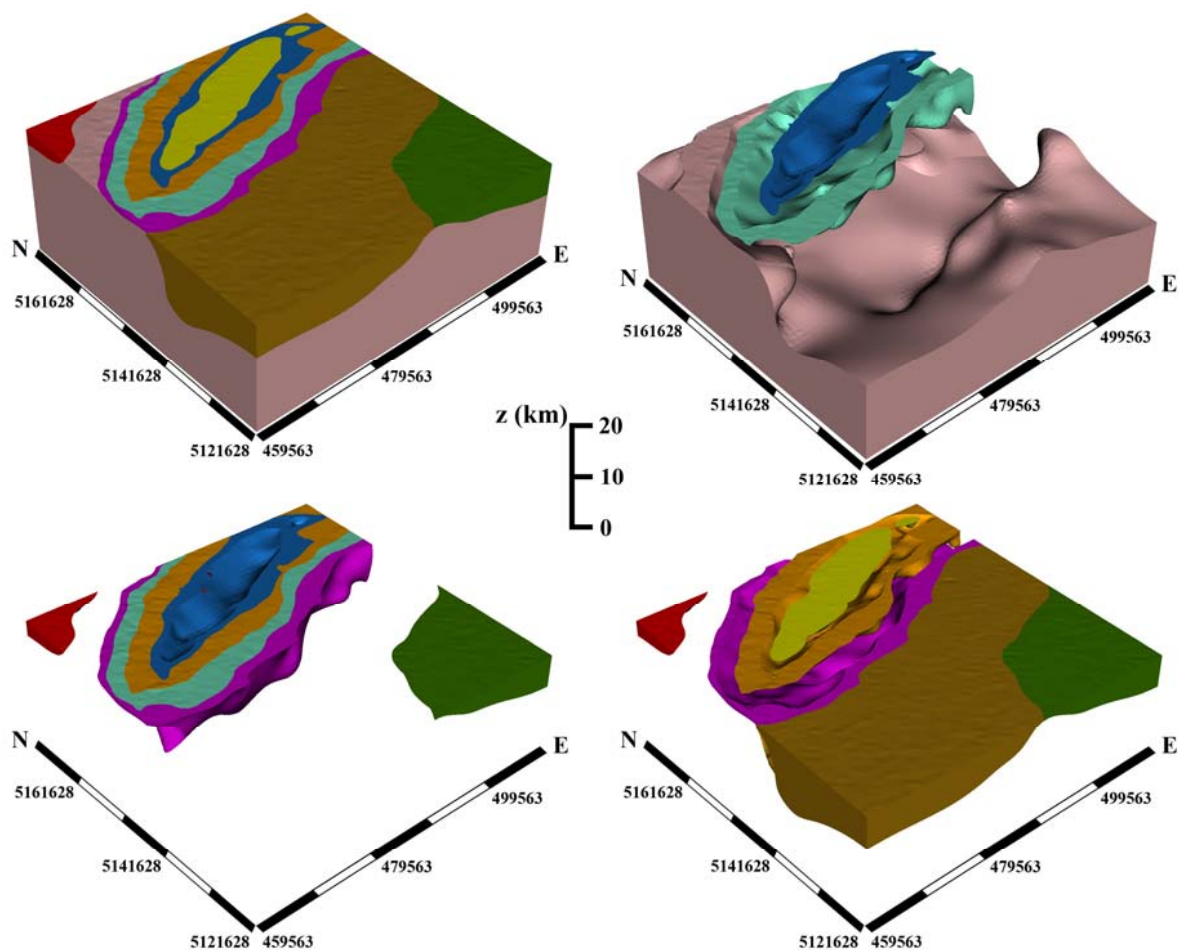

**Figure 5**. Views of the 3D model in GeoModeller. The 3D model takes into account contacts, structural data, and orientation data and follows the order of the stratigraphic succession of geologic units. Color of units is the same as in **Figure 4**. (Cartographic reference system NAD1927 UTM Zone 17N).

### 6.2 Geochemical modeling

Based on the 109 representative outcrop samples we statistically evaluated the abundances of U and Th in the nine units. Analyses of the GT unit and the units representing the SIC were combined with data from the H14 model and from compiled geochemical databases.

For the six units with more than 10 samples (**Table 3**), the distribution function of U and Th concentrations is graphically evaluated using univariate statistics by means of frequency histograms. In order to discriminate the normal and lognormal distributions, the Kolmogorov-Smirnov (K-S) statistical test was applied, providing a p-value for rejecting the null hypothesis. The mean and standard deviation are calculated and used for the geochemical modeling of the other three units (CM, OW, and CT), characterized with less than five samples, corresponding approximately to 1% of the total volume of the CUC.

The first refinement in the geochemical modeling compared to H14 consisted in the use of collected rock samples to describe the chemical composition of the Whitewater Group, a sedimentary and volcanic sequence that fills the Sudbury Basin, as three different lithographic sequences with distinct volumes in the 3D geophysical model (CM, OW, and OP in **Figure 4**). In H14 the Whitewater Group was included with the Huronian Supergroup as a single unit with relatively high U ($4.2^{+2.9}_{-1.7}$ µg/g) and Th ($11.1^{+9.2}_{-4.8}$ µg/g) abundances. In this study, the turbidite wacke of the CM and the siltstone of OW, belonging to the same proximal turbiditic sequence, are characterized as a separate lithographic section with the same average U (~ 1 µg/g) and Th (~ 5 µg/g) abundances (**Table 3**), which are slightly lower than in other sedimentary units and this feature reflects their enrichment in carbonate. The breccia and igneous-textured rocks of the OP are enriched in U and Th with respect to the rest of the Whitewater group and show a normal distribution, with a relative low uncertainty (15%).

The geochemical inputs for modeling the main mass of the SIC come from a combined dataset that includes the samples reported in this study and the compiled database analyzed in H14, i.e. ICPMS compositional data reported in [*Lightfoot et al.*, 1997] (see supporting information **Table S3** and **Table S6**). **Table 2** reports the results of exploratory data analysis considering the two datasets separately and all the data together. The central values of U and Th abundances agree at 1 sigma level with the values reported in [Mareschal et al., 2017]. Although the previous and the new data are characterized by different sources, measurement methodology, and sampling strategies, our analysis demonstrate that the two datasets belong to the same population and can be treated as a single distribution. In the Surrounding Upper Crust (SUC) (see **Figure 1**), the U ($2.0^{+0.4}_{-0.2}$ µg/g) and Th ($10.5^{+1.3}_{-1.1}$ µg/g) abundances associated to the "Sudbury Igneous Complex" unit are obtained by equally weighting the values of the GN and NG units, in agreement with the mixing reported in H14.

**Table 2.** Exploratory data analysis results for U and Th abundance of the GN and NG units which compose the main mass of the SIC (**Figure 4**).

| Dataset | Sudbury Igneous Complex | | | | | |
|---|---|---|---|---|---|---|
| | Granophyre (GN) | | | Norite-gabbro (NG) | | |
| | Number of samples | U ± σ [µg/g] | Th ± σ [µg/g] | Number of samples | U ± σ [µg/g] | Th ± σ [µg/g] |
| H14 | 25 | 3.3 ± 0.2 | 14.9 ± 1.0 | 99 | 1.3 ± 0.4 | $5.9^{+1.9}_{-1.5}$ |
| This study | 10 | 3.4 ± 0.5 | 15.1 ± 2.4 | 9 | 1.3 ± 0.8 | 6.8 ± 4.2 |
| All data | 35 | 3.3 ± 0.3 | 15.0 ± 1.5 | 108 | $1.2^{+0.6}_{-0.4}$ | $5.9^{+2.1}_{-1.5}$ |

The dataset adopted for the geochemical characterization of the Huronian Supergroup and minor felsic and mafic Intrusions unit (HI in **Figure 4**) includes 41 samples belonging to the Huronian Supergroup (Geocode 17b, 18a, 18c, 19a, 19b, 20a, 20b, 21) and the 10 rock samples representative of the minor mafic (Geocode 23d, Nipissing mafic sills) and felsic intrusions (Geocode 30a, Creighton and Murray granite). The frequency histograms and K-S test (**Figure 6**) indicate that the U and Th concentrations in the HI unit are positively skewed and fit a lognormal distribution. The parameters, µ and σ, obtained from the lognormal probability density function (**Figure 6**) give the central tendency and the asymmetrical uncertainties of U and Th abundances (**Table 3**). The U and Th abundances of the HI unit in the CUC are $2.3^{+4.0}_{-1.5}$ µg/g and $8.0^{+15.3}_{-5.3}$ µg/g, respectively. This lower, revised estimate for the Huronian Supergroup, as compared to that reported in H14, results from a targeted and refined collection of samples specifically aimed at the geochemical characterization of the unit. In H14 Huronian Supergroup samples had an anomalous geographical distribution since they were collected only in the western portion of the study area. At the same time, there was an additional lithographic bias as the extensive amount of arkose and quartz arenites in the Mississage Fm. close to SNO+ was not characterized with a proportionate number of samples.

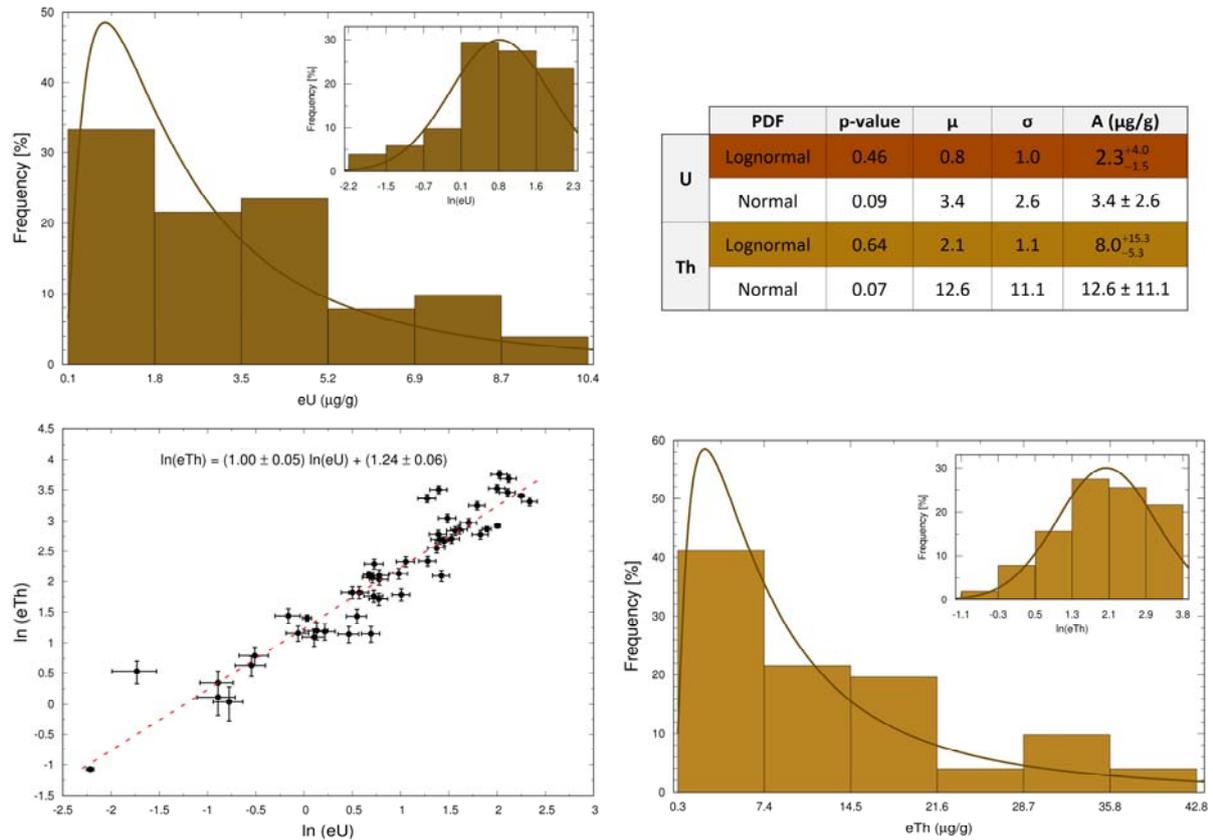

**Figure 6.** Frequency histograms for (top left) U and (bottom right) Th measurements of HI fitted with a lognormal distribution and for the logarithm abundances fitted with a normal distribution. The p-value obtained from the K-S test and the parameters of the fit (μ and σ), considering a lognormal or a normal distribution, are reported in the table on the top right plot together with the result in terms of abundances and uncertainties (A). The plot of the correlation of U and Th abundances and the result of the fit are reported in the bottom left plot, where the error bars refer to experimental uncertainty during measurement.

Although the Gneissic Tonalite suite unit (GT in **Figure 4**), constituted by tonalitic gneiss and minor paragneiss, is only the 4 % of the area of the CUC, it is supposed to be representative of the high-grade gneissic rock of the rest of the upper crust [*Huang et al.*, 2014]. The GT unit has limited exposure (**Table 1**) in the northwest, but comprises 63.7% of the volume of the CUC (**Table 3**). Due to its relevance for estimating the geoneutrino signal, data from the 9 collected samples were integrated with 37 other samples (supporting information **Table S7**) extracted from compiled databases [*Ayer et al.*, 2010; *Beakhouse*, 2011; *Berger*, 2012] on the base of both lithologic and geographic criteria. The final dataset includes the tonalite gneiss samples, attributed to Geocode 11 (gneissic tonalite suite) and Geocode 12 (foliated tonalite suite) of the reference map. The same statistical analysis adopted for the HI unit was applied to the updated

GT unit, which shows a lognormal distribution for U and Th concentrations (**Figure 7**) and agree with the values adopted for the modeling of the "Tonalite/tonalite gneiss" unit in H14 ($0.7^{+0.5}_{-0.3}$ µg/g for U and $3.1^{+2.3}_{-1.3}$ µg/g for Th).

**Table 3.** Summary of geophysical and geochemical and geophysical properties of the units. For each modeled unit the geophysical properties (volume, density and mass) and the U and Th abundances are reported together with the number of samples used for their characterization. The mass uncertainty is obtained by summing the volume uncertainty from the estimation errors of the depth to the top of the middle crust, i.e. 4.7%, and the density uncertainty derived from H14. The correlation coefficient r, with the exception of GN and OP units, is calculated assuming logarithmic distribution of the U and Th abundances.

| Unit | Volume [$10^3$km$^3$] | Volume [%] | Density [g/cm$^3$] | Mass [$10^{15}$kg] | Number of samples | U ± σ [µg/g] | Th ± σ [µg/g] | r |
|---|---|---|---|---|---|---|---|---|
| GT | 29.69 ± 1.40 | 63.7 | 2.73± 0.08 | 81.05± 5.01 | 46 | $0.7^{+1.0}_{-0.4}$ | $2.7^{+6.0}_{-1.9}$ | 0.81 |
| HI | 10.52 ± 0.49 | 22.6 | 2.75± 0.04 | 28.93± 1.79 | 51 | $2.3^{+4.0}_{-1.5}$ | $8.0^{+15.3}_{-5.3}$ | 0.95 |
| NG | 2.64 ± 0.12 | 5.7 | 2.83 ± 0.10 | 7.47± 0.46 | 108 | $1.2^{+0.6}_{-0.4}$ | $5.9^{+2.1}_{-1.6}$ | 0.84 |
| GN | 1.43 ± 0.07 | 3.1 | 2.70± 0.10 | 3.86 ± 0.32 | 35 | 3.3 ± 0.3 | 15.0 ± 1.5 | 0.58 |
| OP | 0.94 ± 0.04 | 2.0 | 2.77± 0.04 | 2.60± 0.22 | 12 | 3.1 ± 0.6 | 8.2 ± 1.0 | -0.15 |
| GF | 0.83 ± 0.04 | 1.8 | 2.73± 0.08 | 2.27± 0.12 | 10 | $2.7^{+3.4}_{-1.5}$ | $10.9^{+17.3}_{-6.7}$ | 0.89 |
| OW | 0.30 ± 0.01 | 0.6 | 2.68± 0.04 | 0.80± 0.05 | 2 | 1.1 ± 0.01 | 5.2 ± 1.5 | - |
| CM | 0.23± 0.01 | 0.5 | 2.75 ± 0.04 | 0.62 ± 0.05 | 4 | 1.1 ± 0.1 | 5.1 ± 0.7 | - |
| CT | 0.04 ± 0.002 | 0.1 | 2.65± 0.02 | 0.11± 0.01 | 2 | 1.8 ± 1.1 | 56.9 ± 27.3 | - |

The composition of the Cartier Granite unit, (CT in **Figure 4**) which is characterized by a poor exposure (**Table 1**) and a relatively small volume in the CUC (**Table 3**), is inferred from the analysis of two samples. The U and Th abundances measured are in agreement with the range reported in Table 1 of [*Meldrum et al.*, 1997] and that for the "Felsic intrusion" unit of H14. These rocks have an anomalous high average Th/U ratio of ~32 compared to average continental crust Th/U = 4.3 [*Rudnick and Gao*, 2003].

The 10 samples from the Grenville Front Tectonic Zone unit (GF in **Figure 4**), occupying the southeast portion of the CUC and corresponding to 1.8% of the total volume, have significant compositional variability (**Table 1**) linked to the different lithologies (gneisses, felsic, mafic, and migmatitic rocks). Results of K-S statistical tests reveal their U and Th abundances and uncertainties are lognormally distributed (**Table 3**).

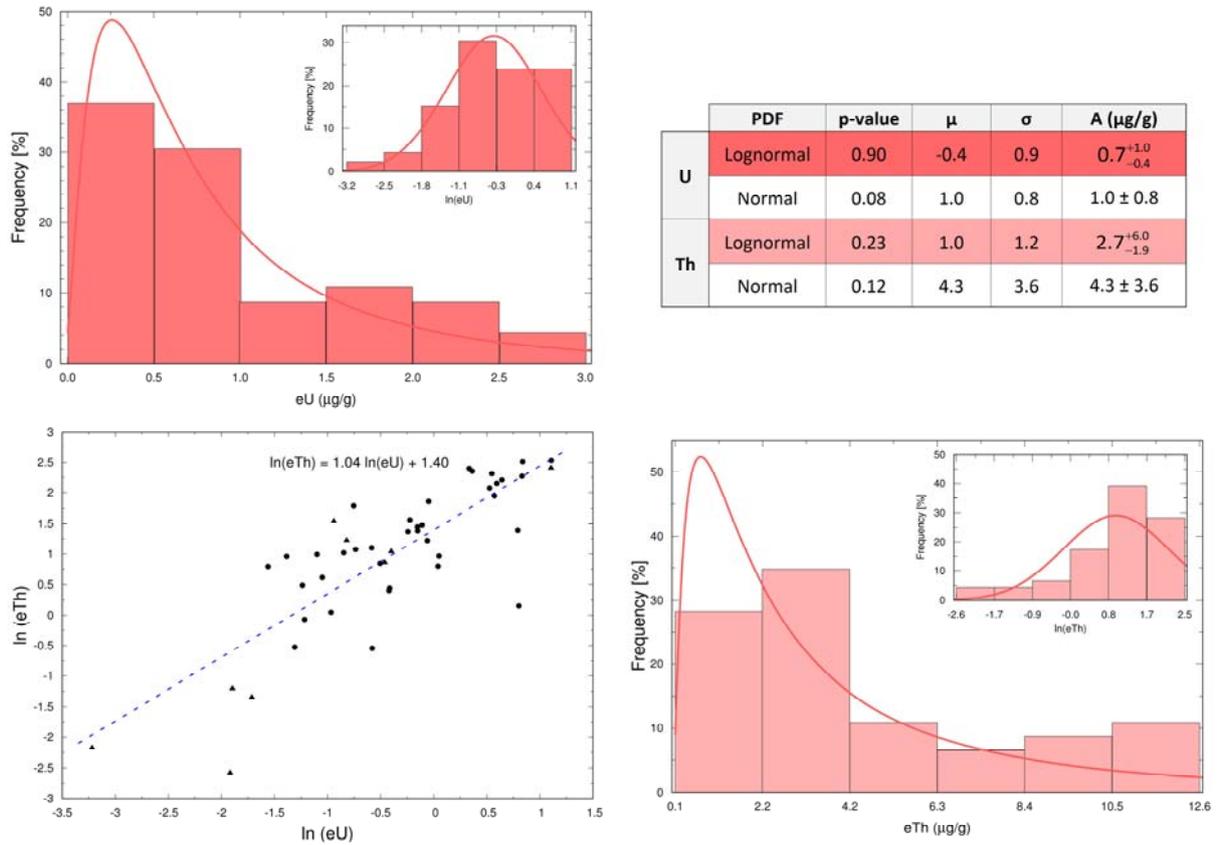

**Figure 7.** Frequency histograms for (top left) U and (bottom right) Th measurements of GT fitted with a lognormal distribution and for the logarithm abundances fitted with a normal distribution. The p-value obtained for the K-S test and the parameters of the fit (μ and σ), considering a lognormal or a normal distribution, are reported in the table on the upper-left panel together with the result in term of abundances (a) and uncertainties. The plot of the correlation of U and Th abundances and the result of the fit are reported in the bottom left plot (triangles refer to samples collected in this study; dots refer to data from compiled databases).

## 7. Geneutrino signal calculation

Predicting a geoneutrino signal at a detector depends upon: (1) the abundance and distribution of Th and U, (2) propagation of the electron antineutrino from the decay point to the detector, and (3) detection of the particle via the Inverse Beta Decay (IBD) reaction within the detector. The final 3D crustal model for the CUC was divided into cells of 0.1 km × 0.1 km × 0.1 km dimensions, for a total of about $5 \times 10^7$ voxels (see supporting information **Dataset S1**). Spatial, geophysical, and geochemical attributes were assigned to each voxel.

The activity of the individual isotopes (i.e., the average number of decays occurring per unit time) for each voxel was computed by dividing the number of radioactive nuclei by the corresponding radioisotope mean lifetime, the former estimated on the base of the radioisotope abundance and the volumetric density defined by the 3D model. The geoneutrino flux reaching SNO+ is then calculated by applying the isotropic $1/4\pi r^2$ spherical scaling factor, weighted for the corresponding geoneutrino spectrum (normalized to the number of geoneutrinos emitted per decay) [*Fiorentini et al.*, 2007], and oscillated by the electron antineutrino three-flavor survival probability [*Capozzi et al.*, 2014] calculated with $\sin^2\theta_{12}= 2.97\cdot10^{-1}$, $\sin^2\theta_{13} = 2.15\cdot10^{-2}$, $\delta m^2 = 7.37\cdot10^{-5}$ eV$^2$, $\Delta m^2 = 2.25 \times 10^{-3}$ eV$^2$ [*Capozzi et al.*, 2017].

Finally, the geoneutrino signal (in TNU) and spectra (**Figure 9**) originating from each cell are calculated combining U and Th oscillated geoneutrino fluxes with IBD cross section. The predicted geoneutrino signals originating by U and Th in the nine units of the CUC are reported in (**Table 4**). The geophysical and geochemical uncertainties associated to each unit are propagated to obtain the geoneutrino signal uncertainties.

**Table 4**. Geoneutrino signals and uncertainties (σ) in TNU for uranium ($S_U$), thorium ($S_{Th}$) and total signals ($S_{TOT}$) for the nine units of the CUC. In the first two columns are reported the geoneutrino signals from U ($G_U$) and Th ($G_{Th}$) calculated with unitary abundances.

| Unit | $G_U \pm \sigma$ | $G_{Th} \pm \sigma$ | $S_U \pm \sigma$ | $S_{Th} \pm \sigma$ | $S_{TOT} \pm \sigma$ |
|---|---|---|---|---|---|
| GT | $0.70 \pm 0.05$ | $0.041 \pm 0.003$ | $0.5^{+0.7}_{-0.3}$ | $0.11^{+0.24}_{-0.07}$ | $0.6^{+0.9}_{-0.4}$ |
| HI | $1.51 \pm 0.09$ | $0.101 \pm 0.006$ | $3.5^{+6.1}_{-2.2}$ | $0.8^{+1.5}_{-0.5}$ | $4.7^{+7.6}_{-2.7}$ |
| NG | $0.72 \pm 0.05$ | $0.049 \pm 0.003$ | $0.9^{+0.4}_{-0.2}$ | $0.29^{+0.09}_{-0.07}$ | $1.2^{+0.4}_{-0.3}$ |
| GN | $0.19 \pm 0.01$ | $[1.25 \pm 0.10] \times 10^{-2}$ | $0.62^{+0.09}_{-0.07}$ | $0.19 \pm 0.02$ | $0.80 \pm 0.08$ |
| OP | $[6.5 \pm 0.4] \times 10^{-2}$ | $[4.2 \pm 0.3] \times 10^{-3}$ | $0.20 \pm 0.04$ | $[3.50 \pm 0.40] \times 10^{-2}$ | $0.24 \pm 0.04$ |
| GF | $[1.4 \pm 0.1] \times 10^{-2}$ | $[7.9 \pm 0.6] \times 10^{-4}$ | $\left[2.6^{+3.5}_{-1.5}\right] \times 10^{-2}$ | $\left[0.9^{+1.4}_{-0.5}\right] \times 10^{-2}$ | $\left[3.6^{+4.5}_{-2.1}\right] \times 10^{-2}$ |
| OW | $[1.55 \pm 0.10] \times 10^{-2}$ | $[9.8 \pm 0.6] \times 10^{-4}$ | $[1.70 \pm 0.02] \times 10^{-2}$ | $[0.51 \pm 0.15] \times 10^{-2}$ | $[2.2 \pm 0.2] \times 10^{-2}$ |
| CM | $[1.24 \pm 0.08] \times 10^{-2}$ | $[7.8 \pm 0.5] \times 10^{-4}$ | $[1.37 \pm 0.15] \times 10^{-2}$ | $[0.40 \pm 0.06] \times 10^{-2}$ | $[1.8 \pm 0.2] \times 10^{-2}$ |
| CT | $[2.03 \pm 0.12] \times 10^{-4}$ | $[1.05 \pm 0.06] \times 10^{-5}$ | $[0.4 \pm 0.2] \times 10^{-3}$ | $[0.6 \pm 0.3] \times 10^{-3}$ | $[1.0 \pm 0.4] \times 10^{-4}$ |

Geochemical uncertainties on the geoneutrino signal were estimated taking into account correlations between U and Th abundances (**Table 3**) and their distributions as follows.

1. For the GT, HI, NG and GF units a bivariate normal distribution describing the joint (ln(U), ln(Th)) Probability Density Function (PDF) was built. For each unit the adopted statistical parameters are the logarithmic U and Th mean and sigma values calculated from the abundances reported in **Table 3**, and the logarithmic U and Th covariance coefficient determined from U and Th concentrations of individual samples.

2. For the GN unit a bivariate normal distribution characterizing the joint (U, Th) PDF was modeled. The statistical parameters are the mean and sigma values reported in **Table 3** for U and Th and the covariance coefficient determined from U and Th concentrations of individual samples.

3. For the OP unit the r = -0.15 correlation coefficient indicates a non evident correlation among U and Th abundances (**Table 3**): the U and Th geochemical distributions are separately modeled as individual normal PDFs having as U and Th mean and sigma the values reported in **Table 3**.

4. For the OW, CM and CT units the number of collected samples is not sufficient for establishing a correlation between U and Th concentrations. For each unit the U and Th geochemical distributions are built as distinct normal PDFs having as U and Th mean and sigma the values reported in **Table 3.**

The mentioned geochemical PDFs together with the geophysical uncertainties are the input ingredients of a Monte Carlo uncertainty propagation procedure: by performing $10^4$ Monte Carlo iterations, the U, Th, and total geoneutrino signal distributions have been built, which are characterized by the median ± 1σ values reported in **Table 4**.

We used the same approach for predicting the geoneutrino signals and their uncertainties for the SUC (**Figure 1**), the Local Middle Crust (LMC), and the Local Lower Crust (LLC) (**Table 5**) which takes into account the geophysical and geochemical inputs reported in Table 4 of H14. The only exception is that of "Sudbury Igneous Complex", which we assigned U and Th abundances on the base of the geochemical considerations described in Section 6.2.

**Table 5.** Summary of geoneutrino signals and uncertainties (σ) in TNU from uranium ($S_U$), thorium ($S_{Th}$), and total signals ($S_{TOT}$) for different components of the LOcal Crust (LOC). Local Upper Crust (LUC), Close Upper Crust (CUC), Surrounding Upper Crust (SUC), Local Middle Crust (LMC), and Local Lower Crust (LLC) are the building blocks defined in **Figure 1** and used for modeling the crust surrounding SNO+.

|   |   | $S_U \pm \sigma$ | $S_{Th} \pm \sigma$ | $S_{TOT} \pm \sigma$ |
|---|---|---|---|---|
| LUC | CUC | $6.1^{+6.2}_{-2.4}$ | $1.6^{+1.7}_{-0.6}$ | $7.7^{+7.7}_{-3.0}$ |
|  | SUC | $4.1^{+1.0}_{-0.7}$ | $1.0^{+0.4}_{-0.3}$ | $5.2^{+1.1}_{-0.8}$ |
| LMC |  | $0.9^{+0.5}_{-0.3}$ | $0.3^{+0.2}_{-0.1}$ | $1.2^{+0.6}_{-0.4}$ |
| LLC |  | $0.4^{+0.3}_{-0.2}$ | $0.2^{+0.2}_{-0.1}$ | $0.6^{+0.4}_{-0.2}$ |
| Total |  | $12.0^{+6.2}_{-2.7}$ | $3.3^{+1.8}_{-0.9}$ | $15.3^{+7.7}_{-3.3}$ |

The calculation of the geoneutrino signal of the Far Field Crust (FFC; **Figure 1**) and Continental Lithospheric Mantle (CLM) (**Table 6**) is described in [*Huang et al.*, 2013] and updated with oscillation parameters from [*Capozzi et al.*, 2017].

**Table 6**. Summary of the total geoneutrino signal and uncertainties (σ) in TNU from the different components of the lithosphere.

|   | $S_{TOT} \pm \sigma$ |
|---|---|
| LOC | $15.3^{+7.7}_{-3.3}$ |
| FFC | $15.2^{+2.7}_{-2.4}$ |
| Bulk Crust | $31.1^{+8.0}_{-4.5}$ |
| CLM | $2.1^{+3.0}_{-1.3}$ |
| Lithosphere | $34.2^{+9.2}_{-5.3}$ |

## 8. Heat production

According to [*Mareschal et al.*, 2017] the CUC is located in a geothermally anomalous region, the Sudbury Structure, with a mean heat flux of 50 ± 7 mW/m$^2$; this flux is larger than the flux typical of the Superior Province of 40 ± 8 mW/m$^2$. The bulk crustal radioactivity has been estimated through inversion of heat flux measurements [*Perry et al.*, 2009], however this approach yields a nonunique constraint for modeling the geoneutrino flux. The energy released by K, Th, and U decay chains provides the crustal radiogenic power, whereas the current geoneutrino detection method (i.e., Inverse Beta Decay reaction) only measures geoneutrinos produced by U and Th decay chains. Estimating the geoneutrino signal from heat flux data requires, among others, the following inputs: (i) the Moho heat flow, (ii) the amount of heat producing elements in the crust, (iii) heat flux data from deep boreholes, and (iv) models that constrain horizontal and vertical heat transport.

Given the U, Th, and K abundances and lithologic densities, one can calculate the corresponding heat production per unit volume, *H*:

$$H(\mu W m^{-3}) = \rho \times \left(0.0985[U] + 0.0263[Th] + 0.0333[K]\right)$$

where concentrations of [*U*] and [*Th*] are in µg/g, and [*K*] is in %, and ρ is density in g/cm$^3$. Adopting the element specific heat generation in µW/g from [*Dye*, 2012], the geochemical abundances in **Table 1** and the densities in **Table 3**, we calculated the *H* values for each Geocode of the geological reference map in the CUC (**Figure 2**).

A heat flux map does not discriminate heat production contributions of U and Th ($H_{U+Th}$) from K ($H_K$) and such maps have an inherent problem with accurately predicting a geoneutrino signal. In typical crustal rocks, contributions to surface heat flux from K heat production can represent up to 30% of the total signal. Uncertainty estimates from $H_K/H$ can vary significantly among different lithologies. The Mississage Fm. of the Huronian Supergroup and the Onaping Formation of the Whitewater Group, which together cover more than 30% of the CUC area (**Table 1**), have $H_K/H \sim 10\%$, whereas the GT unit, which occupies 63.7% by volume of the Close Upper Crust, has a $H_K/H \sim 22\%$. Mafic and ultramafic intrusive rocks of HS and sandstones of Serpent Fm. have $H_K/H \sim 4\%$ and $H_K/H \sim 29\%$ respectively.

Our distribution of *H* values (**Figure 8**) is comparable with that reported in Figure 4 of [*Phaneuf and Mareschal*, 2014]. Even though the study area in [*Phaneuf and Mareschal*, 2014] is wider than the CUC, the histograms of spatial frequency of *H* show comparable lognormal distributions (**Figure 8**) with central values that are compatible at the 1σ level.

Heat production for the Granophyre and Norite Gabbro subunits of the SIC are 2.3 ± 0.3 and 1.0 ± 0.5 µWm$^{-3}$, respectively (**Table 1**), in agreement with that reported in Table 5 of [*Mareschal et al.*, 2017]. The predicted mean heat production of the SIC is 1.6 ± 0.6 µWm$^{-3}$, consistent with it being a melt sheet of upper crustal (high heat production) [*Darling et al.*, 2010] and lower crustal (low heat production) [*Mungall et al.*, 2004] lithologies. The average heat production in the CUC, weighted according to our 3-D model, is $1.0^{+0.8}_{-0.3}$ µWm$^{-3}$. Adding contributions from the Middle and Lower crust yields a total heat production above the Moho of $0.7^{+0.4}_{-0.2}$ µWm$^{-3}$.

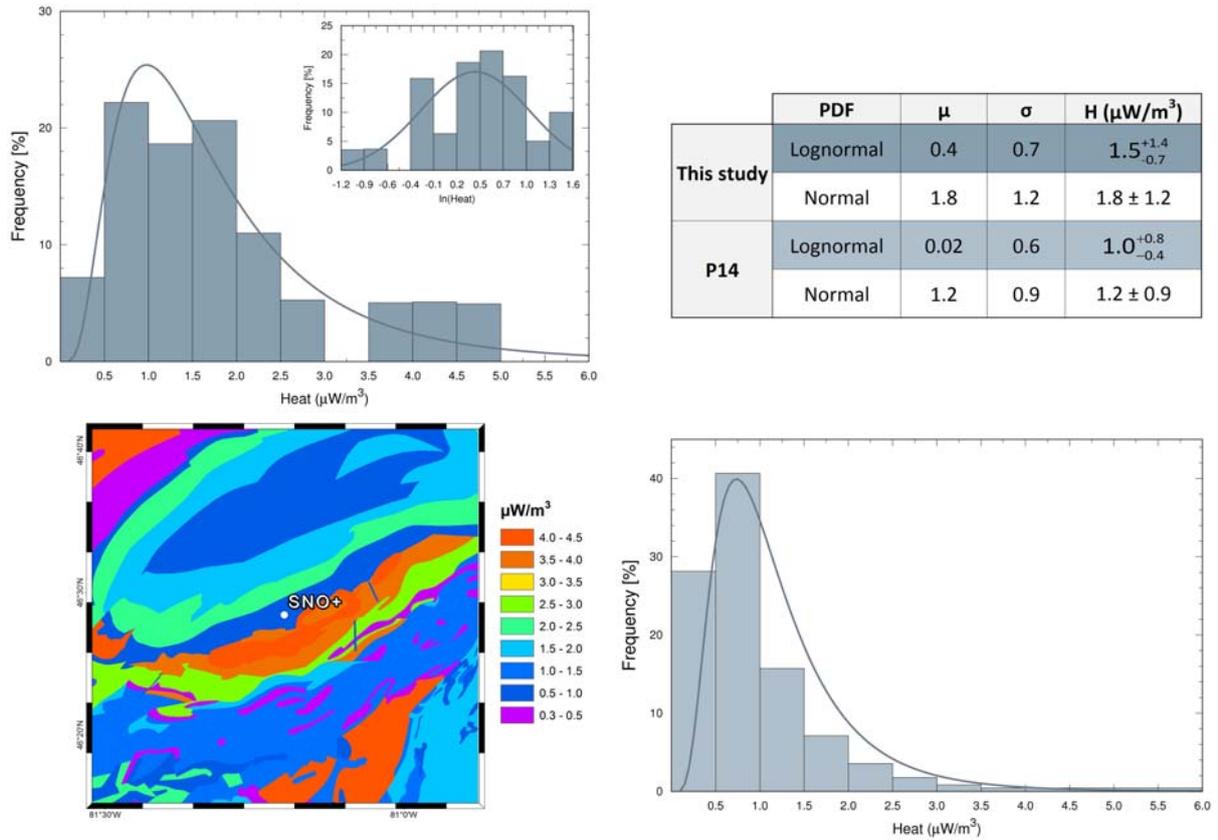

**Figure 8**. Spatial distribution of H values in the CUC is reported in the bottom left plot. In the table are reported the parameters (μ and σ) of the fit considering a lognormal or a normal distribution of the spatial frequencies obtained in this study (top left plot) and in [*Phaneuf and Mareschal*, 2014] (bottom right plot) (P14) together with the results in term of *H* and its uncertainties.

## 9. Discussion

With the aim of improving our model of the geoneutrino flux originating from units surrounding SNO+, we initiated a strategy of dense sampling. The strength of the adopted approach is to prevent a potential bias introduced by compiled literature data [*Huang et al.*, 2014; *Phaneuf and Mareschal*, 2014] that are often motivated by other sampling strategies (e.g. mineral exploration). The Bedrock Geology of Ontario map [*Ontario Geological Survey*, 2011] provided a functional spatial scale for geoneutrino studies in the CUC. This map guided our statistical sampling of units, set the rationale for identifying the independent units, and guided us in establishing the building blocks of the presented model. The sampling resolution (i.e. one sample for 15 km$^2$) was proportional to the surface extent of each cartographic unit.

In H14 the unit including the Huronian Supergroup was predicted to be the dominant near-field, crustal source of the geoneutrino signal at SNO+ and thus it was systematically studied and sampled to improve our knowledge of its composition. The results of the current study highlight the intrinsic heterogeneity of this unit, and the lognormal distribution of U and Th abundances ($2.3^{+4.0}_{-1.5}$ and $8.0^{+15.3}_{-5.3}$ μg/g respectively) and its excellent U-Th correlation (r = 0.95). Any further modeling of the geoneutrino signal at SNO+, following the methodology of this study and H14, will be ineffective without further geophysical characterization of the geochemically heterogeneous Huronian Supergroup. It is a complex mixture of different lithologies that records cyclic deposition during its 200 Ma development toward becoming a passive margin. Glacial events, metamorphic processes, and cross-cutting volcanic fissure-type eruptions have allover-printed this stratigraphic sequence leaving a challenging riddle for the geological community.

In geochemical and environmental surveys, highly incompatible trace elements, such as U and Th, generally follow right skewed distributions: this observation triggered a scientific debate on the a priori adoption of lognormal tendency to describe a statistical population [*Ahrens*, 1954; *Reimann and Filzmoser*, 2000]. The deviation from normality has serious consequences for the statistical treatment of geochemical data since the widespread practice of using the mean and the standard deviation presupposes that data have a Gaussian distribution. In this study, we applied Kolmogorov-Smirnov statistical tests revealing lognormal tendencies of U and Th for the majority of the modeled units (**Table 3**). Where a strong correlation between logarithmic U and Th was observed, a bivariate analysis for the calculation of geoneutrino signal was performed [*Fogli et al.*, 2006], leading to a refinement of the signal uncertainty estimation.

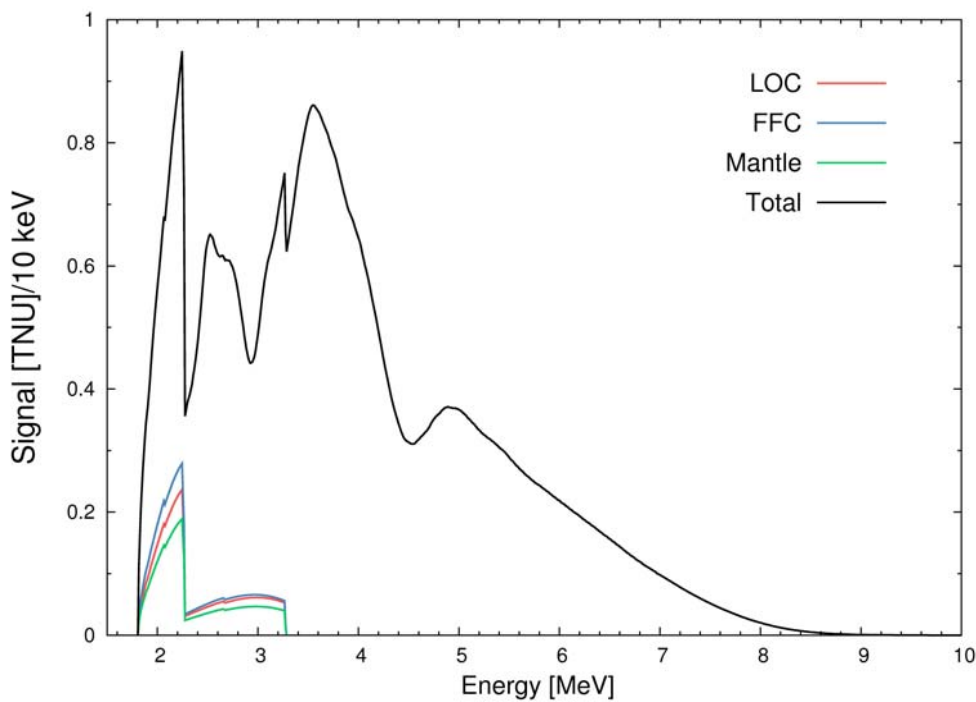

**Figure 9.** Antineutrino spectra expected at SNO+. The geoneutrino spectra are subdivided into the components of LOC (red), FFC (blue), and mantle (green) which includes CLM. The reactor antineutrino spectrum is modeled according to [*Baldoncini et al.*, 2016] and summed to the geoneutrino components to obtain the total antineutrino spectrum (black).

The bulk crust geoneutrino signal expected at SNO+, corresponding to $31.1^{+8.0}_{-4.5}$ TNU, can be expressed as the sum of two comparable and independent contributions, the signal from the 6° × 4° crust surrounding SNO+ (LOC) ($15.3^{+7.7}_{-3.3}$ TNU) and the signal from the rest of global crust (FFC) ($15.2^{+2.7}_{-2.4}$ TNU) (**Table 6**). U and Th in the CUC contributes 51% of the signal (**Table 5**) of the LOC. The signal from the Continental Lithospheric Mantle (CLM) beneath the Mohorovičić discontinuity is calculated according to the model described in [*Huang et al.*, 2013] (**Table 6**).

The overall antineutrino spectrum includes the geoneutrino and the reactor antineutrino components (**Figure 9**), which are modeled according to the predictions discussed in [*Baldoncini et al.*, 2016]. The different portions of geoneutrino spectra contributed by LOC and FFC, particularly in the energy region [1.81 - 2.25 MeV] highlight how differences in Th/U of these two crustal components affect the geoneutrino spectrum expected at SNO+.

The mantle geoneutrino spectrum (**Figure 9**) was built according to a Bulk Silicate Earth (BSE) model constrained by the relative abundances of the refractory lithophile elements in chondritic meteorites [*McDonough and Sun*, 1995], producing a mantle signal of $6.9^{+2.7}_{-2.5}$ TNU.

Competing compositional models for the BSE estimate markedly dissimilar radiogenic power (Q) due to differences in amount of Th and U predicted in the Earth. These estimates were classified [*Dye et al.*, 2015; *Šrámek et al.*, 2013] as low Q, e.g. [*Javoy et al.*, 2010] [*O'Neill and Palme*, 2008] (8 ± 2 TW), medium Q, e.g. [*McDonough and Sun*, 1995] (16.6 ± 3.0 TW), and high Q, e.g. [*Turcotte and Schubert*, 2002] (26 ± 3 TW) models. The estimated mantle geoneutrino signal for low-Q and high-Q models at SNO+ are 3.0 ± 0.7 TNU and $13.5^{+2.6}_{-2.3}$ TNU, respectively. The 1σ uncertainty of geoneutrino signal predicted by LOC encompasses both low and high Q mantle signals, restricting the potential of SNO+ to discriminate between BSE compositional models on the basis of experimental results. On the other hand, by integrating mantle compositional data from Borexino [*Agostini et al.*, 2015] and KamLAND [*Šrámek et al.*, 2016], the results from SNO+ can most usefully be used to resolve U and Th composition of local upper crust belonging to the Southern Province.

**Acknowledgments**

This work was partially funded by the National Institute of Nuclear Physics (INFN) through the ITALian RADioactivity project (ITALRAD) and by the Theoretical Astroparticle Physics (TAsP) research network. The coauthors acknowledge the support of the Geological and Seismic Survey of the Umbria Region (UMBRIARAD), the University of Ferrara (Fondo di Ateneo per la Ricerca scientifica FAR 2016), the Project Agroalimentare Idrointelligente CUP D92I16000030009 and the MIUR (Ministero dell'Istruzione, dell'Università e della Ricerca) under MIUR-PRIN-2012 project. S.A.Wipperfurth gratefully acknowledges support from the support of DOE-INFN exchange program (2016) and the UMD GS Summer Research Fellowship. W.F. McDonough gratefully acknowledges support from NSF EAR 1067983, the University of Maryland, and Tohoku University.

The authors thank Ivan Callegari, Kassandra Raptis, Matteo Albèri, Giovanni Fiorentini, Barbara Ricci, Gerti Xhixha, Enrico Chiarelli, and Carlo Bottardi for useful discussions. Richard Ash for help with the ICPMS analyses. The authors are grateful to Oladele Olaniyan and Richard Smith for providing help and sharing data of the 3D model published in [*Olaniyan et al.*, 2015]. The authors appreciate the essential support and valuable insights from Mike Easton, Mark Chen, Nigel Smith, Eligio Lisi and Pedro Jugo. Finally, we appreciate and thank the careful reviews from J. C. Mareschal and an anonymous reviewer, and insights from and editorial efforts of Uli Faul.

Readers can find detailed geochemistry data and the 3D numerical model in Supporting Information.

**Author contributions**

Virginia Strati, Fabio Mantovani and William McDonough conceived and designed the work as it is. Virginia Strati and Scott Wipperfurth carried out the sampling survey and together with Marica Baldoncini performed the Th and U analyses with HPGe detectors. Scott Wipperfurth carried out the ICPMS measurements. The 3D geophysical model was constructed by Virginia Strati and Fabio Mantovani. All the authors participated in the data analysis and interpretation of the results, while Virginia Strati, Marica Baldoncini, Fabio Mantovani conducted the geoneutrino signal calculation. Virginia Strati, Fabio Mantovani took the lead in designing and composing the manuscript with the input from all the authors. All the authors critically revised and provided final approval of the version submitted.